\documentclass[preprint,aps,jcp,onecolumn,superscriptaddress]{revtex4-2}

\usepackage[paperheight=11in,paperwidth=8.25in,margin=0.67in]{geometry}

\usepackage[T1]{fontenc} 
\usepackage[version=3]{mhchem} 
\usepackage{mathpazo}

\usepackage{amsfonts}
\usepackage{amsmath}
\usepackage{booktabs}
\usepackage{tikz}
\usepackage{threeparttable}
\usepackage{multirow}
\usepackage{appendix}
\usepackage{amssymb}
\usepackage{graphicx}
\usepackage{subfigure}
\usepackage{enumerate}
\usepackage{colortbl}
\usepackage{framed}
\usepackage{verbatim}
\usepackage{amsbsy}
\usepackage{bm}

\usepackage[normalem]{ulem}
\usepackage{float}
\usepackage[linesnumbered,boxed]{algorithm2e}
\usepackage{algorithm2e}

\usepackage{simplewick} 
\usepackage{bbm}

\newcommand{\ii}{\mathbbm{i}}
\newcommand{\sgn}{ {\rm{sgn}} }

\newfloat{Algorithm}{htbp}{alg}
\floatname{Algorithm}{Algorithm}

\newcounter{exe}[figure]
\newcommand{\iexe}{\refstepcounter{exe}\the\value{exe}:}

\begin{document}

\title{On the bottom-up construction of many-electron relativistic QED Hamiltonian}

\author{Wenjian Liu}\email{liuwj@sdu.edu.cn}
\affiliation{Qingdao Institute for Theoretical and Computational Sciences and
Center for Optical Research and Engineering, Shandong University, Qingdao, Shandong 266237, China}

\begin{abstract}
It was shown more than a decade ago [J. Chem. Phys. 139, 014108 (2013)] that a many-electron relativistic quantum electrodynamics
(QED) Hamiltonian for high-precision electronic structure calculations
can be constructed in a bottom-up fashion,
by virtue of charge-conjugated contraction (CCC) of fermion operators when normal-ordering 
the starting unbounded relativistic Hamiltonian (second-quantized in terms of the electronic Dirac field)
with respect to the filled negative-energy Dirac sea of electrons. 
It is shown here that the same relativistic QED Hamiltonian can also be obtained 
by equal average of the two relativistic Hamiltonians 
resulting from the normal-ordering of the starting unbounded relativistic Hamiltonians 
(second-quantized in terms of the electronic and positronic Dirac fields, respectively) with respect 
to the filled negative-energy Dirac seas of electrons and positrons, respectively, via the standard contraction of fermion operators.
In essence, both procedures incorporate properly the fundamental charge-conjugation symmetry of relativistic quantum mechanics 
to ensure the symmetric treatment of the electronic and positronic degrees of freedom. 
\end{abstract}

\maketitle

\section{Introduction}
Quantum electrodynamics (QED) is the highest level of theory for describing electromagnetic interactions between charged particles
and has reached unprecedented accuracy for few-body systems. However, the situation is very different for bound states of many-body
systems, to which rigorous QED cannot be applied, due to huge complexity and tremendous computational cost.  
This is because QED is a time-dependent perturbation theory, where the Wick expansion of 
time-ordered products of interactions at different spacetime points generates an overwhelming number of Feynman diagrams.
For an $N$-electron system, the number of distinct Feynman diagrams 
scales asymptotically as $N!$, not to mention the complicated regularization and renormalization procedures.
As a matter of fact,  the ``first relativity and QED then correlation'' paradigm underlying relativistic QED is intrinsically 
unsuitable for many-electron systems, where  QED effect is generally smaller than relativistic and correlation effects.
Therefore, it is the ``first relativity and correlation then QED'' paradigm\cite{eQED} that should be adopted instead.
To this end, a time-independent many-electron relativistic QED Hamiltonian
incorporating infinite-order relativistic effect and leading-order QED effect should first be introduced and then combined with a wave function Ansatz
for correlation. Such a Hamiltonian can in principle be introduced 
in two ways: top-down reduction or bottom-up construction. The former tries to extract effect QED interaction operators
from the corresponding QED energy expressions, whereas the latter tries to build up a relativistic QED Hamiltonian based solely on the principles
of relativistic quantum mechanics. Following the former procedure, Shabaev\cite{np-eQED} proposed a relativistic QED Hamiltonian yet
under the no-pair approximation (NPA). Moreover, his Hamiltonian is defined only on an active space, which cannot be set up uniquely.
In contrast, the relativistic QED Hamiltonian constructed by us in a bottom-up manner\cite{eQED} is defined on the full Fock space without the NPA.
Interestingly, the same relativistic QED Hamiltonian can be obtained both algebraically\cite{eQED} and diagrammatically\cite{IJQCeQED}.
In particular, the latter clearly demonstrates why the previous formulation\cite{KutzFS} is incomplete.
Although the bottom-up procedure for constructing a complete relativistic QED Hamiltonian
is fully scrutinized\cite{IJQCrelH,PhysRep}, it is still often misunderstood\cite{JapQED,RespCommentQED,CommentQED}.
This is probably because the charge-conjugated second quantization\cite{eQED}, dictated by
the charge-conjugation symmetry of relativistic quantum mechanics, is not documented anywhere else  in
the context of second quantization. In short, the charge-conjugated contraction (CCC)\cite{eQED} of fermion operators
must be adopted when normal-ordering the starting unbounded relativistic Hamiltonian (second-quantized in terms of the electronic Dirac field)
with respect to the filled negative-energy Dirac sea. 
It is shown here that the same relativistic QED Hamiltonian can also be obtained 
by equal average of the two relativistic Hamiltonians 
resulting from the normal-ordering of the starting unbounded relativistic Hamiltonians  
(second-quantized in terms of the electronic and positronic Dirac fields, respectively) with respect 
to the respective filled negative-energy Dirac seas of electrons and positrons, via the standard contraction of fermion operators.
As a result, the CCC emerges more naturally. 

%

The Einstein summation over repeated indices is to be employed throughout.
\section{Bottom-up construction of relativistic QED Hamiltonian}
\subsection{Hamiltonian quantized with Dirac field}
It should be noted in the first place that any second quantization must
start with a definition of the fermion field operator $\hat{\phi}$ in terms of
a vacuum and corresponding annihilation and creation operators, such that the resulting
Hamiltonian takes the whatsoever vacuum as its ground state of zero energy. For instance,
if we start with
\begin{align}
\hat{\phi}&=a_p\psi_p,\quad
a_p|\mathrm{vac}\rangle=0,\quad p\in\mbox{ PES, NES}, \label{vac0}
\end{align}
where the spinors $\{\psi_p\}$ refer to the positive- (PES) and negative-energy (NES) solutions/states
of the Dirac equation
\begin{align}
D\psi_p&=\epsilon_p\psi_p,\label{DEQ}\\
D&=D_0+q\phi_{\mathrm{ext}}(\boldsymbol{r}),\quad q=-1,\\
D_0&=c\boldsymbol{\alpha}\cdot\boldsymbol{p}+\beta mc^2,
\end{align}
we will have the following many-electron Hamiltonian
\begin{align}
\mathcal{H}&=D_p^qa^p_q +\frac{1}{2}g_{pr}^{qs}a^{pr}_{qs},\quad p, q, r, s \in \mbox{PES, NES},\label{Hbase}\\
a^p_q&=a_p^\dag a_q,\quad a^{pq}_{rs}=a_p^\dag a_q^\dag a_s a_r,\\
D_p^q&=\langle\psi_p|D|\psi_q\rangle,\quad
g_{pr}^{qs}=\langle\psi_p\psi_q|g(1,2)|\psi_r\psi_s\rangle.
\end{align}
Here, $D_0$, $\phi_{\mathrm{ext}}$, and $g(1,2)$ are the free-particle Dirac operator, nuclear attraction, and two-body interaction, respectively.
The specific form of $g(1,2)$ is immaterial in the present context. It is just that it must encompass the frequency-dependent Breit interaction
when going beyond the NPA. 
Note that the Hamiltonian \eqref{Hbase}
is already normal ordered with respect to $|\mathrm{vac}\rangle$. However, this Hamiltonian is unbounded and does not distinguish the empty from the
filled Dirac picture. Since the empty Dirac picture must be abandoned (for it implies that no atom would be stable),
we ought to incorporate the filled Dirac picture, by introducing a reference $|0;\tilde{N}\rangle$ ($=\Pi_{\tilde{i}}^{\tilde{N}} a^{\tilde{i}}|0;\tilde{0}\rangle$)
built up with zero PESs and $\tilde{N}$
($\rightarrow\infty$) NESs. The Hamiltonian \eqref{Hbase} must then be normal ordered with respect to
$|0;\tilde{N}\rangle$, 
so as to obtain a physical Hamiltonian $H_n^{\mathrm{QED}}$
\begin{align}
\mathcal{H}&=H_n^{\mathrm{QED}} + C_n,\label{H2}\\
H_n^{\mathrm{QED}}&=H^{\mathrm{FS}}_n + Q_p^q\{a^p_q\}_n,\quad p, q, r, s \in \mbox{PES, NES},\label{HnH}\\
H^{\mathrm{FS}}_n&=D_p^q\{a^p_q\}_n +\frac{1}{2}g_{pr}^{qs}\{a^{pr}_{qs}\}_n,\label{KZFS}\\
Q_p^q&=\bar{g}^{qs}_{pr} \acontraction[0.5ex]{}{a^r}{}{a_s}a^ra_s,\quad \bar{g}_{pr}^{qs}=g^{qs}_{pr}-g^{sq}_{pr},\label{Qdef}\\
C_n&=\langle 0;\tilde{N}|\mathcal{H}|0;\tilde{N}\rangle \\
&=D_p^{q}\acontraction[0.5ex]{}{a^p}{}{a_q}a^pa_q+\frac{1}{2}\bar{g}_{pr}^{qs}\acontraction[0.5ex]{}{a^p}{}{a_q}a^pa_q \acontraction[0.5ex]{}{a^r}{}{a_s}a^ra_s.\label{HnC}
\end{align}
The question is how to perform the contractions. Three types of contractions
were examined by Inoue and coworkers\cite{JapQED}:
\begin{align}
\acontraction[0.5ex]{}{a^p}{}{a_q}a^pa_q&=
\begin{cases}
0 \quad \mbox{ (constantly null contraction (CNC))};\\
\langle 0,\tilde{N}|a^p_q|0,\tilde{N}\rangle \quad \mbox{ (conventional contraction (CC))};\\
\frac{1}{2}\langle 0,\tilde{N}|[a^p, a_q]|0,\tilde{N}\rangle \quad \mbox{ (charge-conjugated contraction (CCC))}.\label{CCC}
\end{cases}
\end{align}
It is obvious that the CNC gives rise to a zero one-body potential $Q$ \eqref{Qdef}, so that $H_n^{\mathrm{QED}}$ \eqref{HnH} is just the Fock space Hamiltonian
$H^{\mathrm{FS}}_n$ \eqref{KZFS} advocated by Kutzlnigg, which misses \emph{genuine} QED effects (vacuum polarization (VP) and electron self-energy (ESE), etc.)\cite{eQED}.
As for the CC, one readily obtains 
\begin{align}
Q_p^q&=\bar{g}^{q\tilde{i}}_{p\tilde{i}},\quad \tilde{i}\in\mbox{ NES},\label{CCQpot}
\end{align}
by virtue of the elementary anticommutation relations (ACR) of fermion operators
\begin{subequations}\label{Commut}
\begin{equation}
a^p a^q + a^q a^p=0,
\end{equation}
\begin{equation}
a_pa_q + a_qa_p=0,
\end{equation}
\begin{equation}
a^p a_q + a_q a^p=\delta^p_q.\label{Fermi}
\end{equation}
\end{subequations}
It is obvious that the so-obtained Hartree-Fock-like potential \eqref{CCQpot} is infinitely repulsive due to the existence of infinitely
many negative-energy electrons. In other words, no atom would be stable under the filled Dirac picture, just like the empty Dirac picture.
In contrast, the CCC gives rise to the following effective one-body potential\cite{eQED}
\begin{align}
Q_p^q&=-\frac{1}{2}\bar{g}_{ps}^{qs}\sgn({\epsilon_s}).\label{CCCQpot}
\end{align}
Different from the expression \eqref{CCQpot}, the $Q$ potential \eqref{CCCQpot} involves all PES and NES,
whether occupied or not. As a matter of fact, 
the direct and exchange terms of $Q$ correspond\cite{PhysRep} to precisely the VP and ESE, respectively,
although a suitable regularization and renormalization should first be carried out before it
can actually be evaluated. The CCC was introduced\cite{eQED} \emph{a priori} to incorporate directly the charge-conjugation symmetry,
so as to treat electrons and positrons on an equal footing. An alternative manipulation is provided in the next section.

\section{Hamiltonian quantized with charge-conjugated Dirac fields}\label{CCC-H}
The charge conjugation is defined as
\begin{equation}
\hat{C}=\mathbf{C}_0\hat{K}_0,\quad \hat{C}^\dag=\hat{C}^{-1}=\hat{C},
\quad \mathbf{C}_0=-\ii \alpha_y\beta=\begin{pmatrix}\mathbf{0}_2&i\sigma_y\\
-i\sigma_y&\mathbf{0}_2\end{pmatrix}=\mathbf{C}_0^\dag=\mathbf{C}_0^{-1},\label{CCsymm}
\end{equation}
which transforms the electronic Dirac equation \eqref{DEQ}
to that of a positron (of charge $-q=1$)
\begin{align}
D^C \psi^C_p &=\epsilon_p^C \psi^C_p,\label{PositronDEQ}\\
D^C&=-\hat{C}D\hat{C}^{-1}
=-\mathbf{C}_0^\dag D^* \mathbf{C}_0=D_0-q\phi_{\mathrm{ext}}(\boldsymbol{r}),\label{hDcop} \\
\psi_p^C&=\hat{C}\psi_p=\mathbf{C}_0\psi^*_p,\quad \epsilon_p^C=-\epsilon_p.
\end{align}
Likewise, the four-component Dirac field operator $\hat{\phi}(\boldsymbol{r})$ (in the particle-hole picture)
\begin{align}
\hat{\phi}_{\sigma}(\boldsymbol{r})&=b_p\psi_{p\sigma}(\boldsymbol{r})+b^{\tilde{p}}\psi_{\tilde{p}\sigma}(\boldsymbol{r}),
\quad p\in\mbox{ PES}, \quad \tilde{p}\in\mbox{ NES},\quad \sigma\in[1,4]\label{ElectronField}
\end{align}
will be transformed by charge conjugation to
\begin{align}
\hat{\phi}^C(\boldsymbol{r})&=\mathbf{C}_0 \hat{\phi}^{\dag T}(\boldsymbol{r}) \\
&=b_{\tilde{p}}\psi^C_{\tilde{p}}(\boldsymbol{r})+b^p\psi_p^C(\boldsymbol{r}),\label{PositronField}
\end{align}
where the first and second terms annihilate a positive-energy positron (NB: $\epsilon_{\tilde{p}}^C>0$) and create a positive-energy electron, respectively,
in accordance with the fact that the first and second terms of $\hat{\phi}(\boldsymbol{r})$ \eqref{ElectronField}
annihilate a positive-energy electron and create a positive-energy positron, respectively. That is,
$\hat{\phi}^C(\boldsymbol{r})$ \eqref{PositronField} would be the starting quantized Dirac field, had we lived in the world of antiparticles.

Both $\hat{\phi}(\boldsymbol{r})$ \eqref{ElectronField} and
$\hat{\phi}^C(\boldsymbol{r})$ \eqref{PositronField} are associated with the vacuum $|\mathrm{vac}\rangle=|0;\tilde{0}\rangle$.
To expedite algebraic manipulations, we can rewrite them as\cite{LiuQED2020}
\begin{align}
\hat{\phi}(\boldsymbol{r})&=a_p\psi_p(\boldsymbol{r})+a_{\tilde{p}}\psi_{\tilde{p}}(\boldsymbol{r}),
\quad \epsilon_p>0, \quad \epsilon_{\tilde{p}}<0,\label{ElectronFielda}\\
\hat{\phi}^C(\boldsymbol{r}) &=a_{\tilde{p}}\psi^C_{\tilde{p}}(\boldsymbol{r})+a_p\psi_p^C(\boldsymbol{r}),
\quad \epsilon_{\tilde{p}}^C>0,\quad \epsilon_p^C<0,\label{PositronFielda}
\end{align}
by taking $|0;\tilde{N}\rangle$ (Dirac sea of electrons)
and $|0_{e^+};\tilde{N}_{e^+}\rangle$ (Dirac sea of positrons)
as the vacua, respectively. Eq. \eqref{ElectronFielda} is just Eq. \eqref{vac0}, provided that
$|\mathrm{vac}\rangle$ in the former is chosen to be $|0;\tilde{N}\rangle$.

With the above background, we can calculate
\begin{align}
h^C&=\int \hat{\phi}^{C\dag} D^C \hat{\phi}^C d\tau=\int (\mathbf{C}_0 \hat{\phi}^{\dag T})^\dag D^C (\mathbf{C}_0 \hat{\phi}^{\dag T}) d\tau
=\int \hat{\phi}^{T} (\mathbf{C}_0^\dag D^C \mathbf{C}_0) \hat{\phi}^{\dag T}d\tau\nonumber\\
&=-\int \hat{\phi}^{T} D^* \hat{\phi}^{\dag T}d\tau=-\int \hat{\phi}^{T} D^T \hat{\phi}^{\dag T}d\tau\nonumber\\
&=-\int \hat{\phi}_{\rho} D_{\sigma\rho}\hat{\phi}^{\dag}_{\sigma}=-\int D \hat{\phi}\hat{\phi}^{\dag}d\tau\nonumber\\
&=\int \{\hat{\phi}^{\dag}D \hat{\phi}\}_nd\tau-\int \langle\mathrm{vac}|D\hat{\phi}\hat{\phi}^\dag|\mathrm{vac}\rangle  d\tau\label{hCpositron}\\
&=h_n - D_p^q \langle 0;\tilde{N}|a_q a^p|0;\tilde{N}\rangle,\quad p,q\in\mbox{ PES, NES},\label{hCpositrona}\\
h_n&=D_p^q\{a^p_q\}_n,\label{hnOp}
\end{align}
which is to be compared to the usual second-quantized Dirac operator (cf. the CC in Eq. \eqref{CCC})
\begin{align}
\tilde{h}&=\int \hat{\phi}^{\dag} D \hat{\phi} d\tau=\int \{\hat{\phi}^{\dag}D \hat{\phi}\}_n d\tau
+ \int \langle\mathrm{vac}|\hat{\phi}^\dag D\hat{\phi}|\mathrm{vac}\rangle d\tau\label{helectron}\\
&=h_n + D_p^q \langle 0;\tilde{N}|a^p a_q|0;\tilde{N}\rangle,\quad p,q\in\mbox{ PES, NES}.\label{helectrona}
\end{align}
Since charge conjugation is an inherent symmetry, $\tilde{h}$ \eqref{helectron}/\eqref{helectrona} and
$h^C$ \eqref{hCpositron}/\eqref{hCpositrona} should be averaged with an equal weight, leading to
\begin{align}
h&= \frac{1}{2}(\tilde{h}+h^C)
=\int \{\hat{\phi}^{\dag}D \hat{\phi}\}_n d\tau +\frac{1}{2}\int \langle\mathrm{vac}|[\hat{\phi}^\dag, D\hat{\phi}]|\mathrm{vac}\rangle d\tau\\
&=h_n + C_{n1},\\
C_{n1}&= D_p^q \acontraction[0.5ex]{}{a^p}{}{a_q}a^pa_q, \quad \acontraction[0.5ex]{}{a^p}{}{a_q}a^pa_q=\frac{1}{2}\langle 0;\tilde{N}|[a^p, a_q]|0;\tilde{N}\rangle
=-\frac{1}{2}\delta^p_q \sgn(\epsilon_p),
\end{align}
where $\acontraction[0.5ex]{}{a^p}{}{a_q}a^pa_q$ is just the CCC \eqref{CCC} and arises here naturally from the averaging process.

The two-body operator can be calculated in the same way, viz.,
\begin{align}
G^C &= \frac{1}{2}\int\int \hat{\phi}^{C\dag}(1) \hat{\phi}^{C\dag}(2) V(1,2) \hat{\phi}^{C}(2)\hat{\phi}^{C}(1) d\tau_1 d\tau_2\nonumber\\
&=\frac{1}{2}\int\int \hat{\phi}^T(1)\hat{\phi}^T(2)[\mathbf{C}_0^\dag(1)\mathbf{C}_0^\dag(2)V(1,2)\mathbf{C}_0(2)\mathbf{C}_0(1)]
\hat{\phi}^{\dag T}(2)\hat{\phi}^{\dag T}(1)\nonumber\\
&=\frac{1}{2}\int\int \hat{\phi}^T(1)\hat{\phi}^T(2)V^*(1,2)
\hat{\phi}^{\dag T}(2)\hat{\phi}^{\dag T}(1)\nonumber\\
&=\frac{1}{2}\int\int\hat{\phi}_{\rho}(1)\hat{\phi}_{\sigma}(2)[V(1,2)]_{\delta\rho,\gamma\sigma  }\hat{\phi}^\dag_{\gamma}(2)\hat{\phi}^\dag_{\delta}(1)\nonumber\\
&=\frac{1}{2}\int\int V(1,2)\hat{\phi}(1)\hat{\phi}(2)\hat{\phi}^\dag (2)\hat{\phi}^\dag (1)\\
&=\frac{1}{2}g_{pr}^{qs} a_qa_s a^r a^p\\
&=\frac{1}{2}g_{pr}^{qs}\{a^{pr}_{qs}\}_n -\bar{g}_{pr}^{qs}\langle 0;\tilde{N}|a_sa^r|0;\tilde{N}\rangle\{a^p_q\}_n
+\frac{1}{2}\bar{g}_{pr}^{qs}\langle 0;\tilde{N}|a_qa^p|0;\tilde{N}\rangle\langle 0;\tilde{N}|a_sa^r|0;\tilde{N}\rangle,\label{GCnew}
\end{align}
which should be compared to the usual two-body operator
\begin{align}
\tilde{G}&= \frac{1}{2}\int\int \hat{\phi}^{\dag}(1) \hat{\phi}^{\dag}(2) V(1,2) \hat{\phi}(2)\hat{\phi}(1) d\tau_1 d\tau_2\nonumber\\
&=\frac{1}{2}g_{pr}^{qs} a^{pr}_{qs}\\
&=\frac{1}{2}g_{pr}^{qs}\{a^{pr}_{qs}\}_n+\bar{g}_{pr}^{qs}\langle 0;\tilde{N}|a^ra_s|0;\tilde{N}\rangle\{a^p_q\}_n
+\frac{1}{2}\bar{g}_{pr}^{qs}\langle 0;\tilde{N}|a^pa_q|0;\tilde{N}\rangle\langle 0;\tilde{N}|a^ra_s|0;\tilde{N}\rangle. \label{Gnew}
\end{align}
The average of $\tilde{G}$ and $G^C$ leads to
\begin{align}
G&=\frac{1}{2}(\tilde{G}+G^C)=G_n+C_{n2},\\
G_n&=\frac{1}{2}g_{pr}^{qs}\{a^{pr}_{qs}\}_n + Q_p^q \{a^p_q\}_n,\quad Q_p^q =\bar{g}_{pr}^{qs} \acontraction[0.5ex]{}{a^r}{}{a_s}a^ra_s,
\quad \quad \acontraction[0.5ex]{}{a^r}{}{a_s}a^ra_s=\frac{1}{2}\langle 0;\tilde{N}|a^r, a_s|0;\tilde{N}\rangle,\label{GnOp}\\
C_{n2}&=\frac{1}{2}\bar{g}_{pr}^{qs} [\langle 0;\tilde{N}|a^pa_q|0;\tilde{N}\rangle\langle 0;\tilde{N}|a^ra_s|0;\tilde{N}\rangle
+  \langle 0;\tilde{N}|a_pa^q|0;\tilde{N}\rangle\langle 0;\tilde{N}|a_sa^r|0;\tilde{N}\rangle]. \label{Cn2}
\end{align}
It can be seen that the sum of $h_n$ \eqref{hnOp} and $G_n$ \eqref{GnOp} is just $H_n^{\mathrm{QED}}$ \eqref{HnH} along with the CCC. In particular,
in sharp contrast to the $Q$-potential in Eq. \eqref{GnOp}/\eqref{CCCQpot},
the second term of $G^C$ \eqref{GCnew} or that of $\tilde{G}$ \eqref{Gnew} (see also Eq. \eqref{CCQpot}) is divergent and cannot be rgularized/renormalized.
As such, the averaging of the Hamiltonian $\tilde{h}+\tilde{G}$ for electrons and the Hamiltonian $h^C+G^C$ for positrons (moving in the
same external field $\phi_{\mathrm{ext}}(\boldsymbol{r})$ as electrons)
is a must rather than merely a formal step. Note in passing that
the constant $C_{n2}$ \eqref{Cn2} is somewhat different from the second term of Eq. \eqref{HnC}. The latter is more symmetric due to
the direct use of the CCC. However, such difference does not matter at all, for such constants will be renormalized away.

The above manipulation reveals that the relativistic QED Hamiltonian $H_n^{\mathrm{QED}}$ \eqref{HnH}, especially the Q-potential in Eq. \eqref{CCCQpot}/\eqref{GnOp},
does stem from the symmetric treatment of electrons and positrons, as dictated by the charge conjugation symmetry of relativistic quantum mechanics.
Since any energy-independent Hamiltonian must be linear in the (at most) two-body interaction, it is clear that the CCC-based relativistic QED Hamiltonian $H_n^{\mathrm{QED}}$ \eqref{HnH} is the most accurate many-electron Hamiltonian. Because of this, it serves as the basis of the emerging field of ``molecular QED''\cite{PhysRep}.
Note in passing that, while the $Q$ potential (of $\mathcal{O}(Z^3\alpha^3)$) can nowadays be evaluated directly with a Gaussian basis
\cite{SaueESE2205,SaueVP-WK2026}, the use of short-ranged model operators\cite{Uehling1935,Pyykko-SE2003,FG-SE2005,DyallModelQ,ShabaevModelSE,ShabaevModelSEcode2018,Malyshev2022model}
is more practical for molecular QED calculations\cite{MO-1eQED1,MO-1eQED2,MO-1eQED3,MOQED-JCP2021,Saue4CQED2022}.

As a final point, it might be noticed\cite{RespCommentQED} that, at variance with the CC,
both the CNC and CCC do not satisfy the basic ACR of fermion operators \eqref{Commut}.
However, this does not imply that the physically incorrect CC is good, whereas the physically correct CNC/CCC is bad!
The CNC just ignores\cite{KutzFS} the VP and ESE from the outset, whereas the CCC arises from an averaging process as shown above.
Apart from this, all steps, e.g., Eqs. \eqref{hCpositrona}, \eqref{helectrona}, \eqref{GCnew}, and \eqref{Gnew}, follow strictly the ACR \eqref{Commut}.
As already scrutinized before\cite{LiuQED2020},
the introduction of a filled Dirac sea of electrons (which is not part of the Dirac equation itself) is a must but
not yet complete. By virtue of the charge-conjugation symmetry (which is indeed a property of the Dirac equation),
there exists also a filled Dirac sea of positrons (which is again not part of the Dirac equation itself).
The two seas are coexistent and equivalent, and should hence be averaged with an equal weight of one half,
thereby leading naturally to the CCC\cite{eQED}. Just like that the ACR is dictated by the fermi statistics
of nonrelativistic or relativistic fermions, the CCC is dictated by the charge-conjugation symmetry
of relativistic fermions. Both are fundamental laws of quantum mechanics that
have to be imposed from the outside of quantum mechanical equations, so as to render the latter physically correct
(recalling that the Schr\"odinger equation can also describe bosons, provided that boson statistics is mposed).
Note in passing that the CCC can also be viewed as the time-independent analog
of the symmetric-in-time, equal-time contraction (ETC) of time-dependent fermion operators\cite{Schwinger1951}, which
is embodied automatically in the Feynman fermion propagator\cite{IJQCeQED}, again not merely a formal ingredient.
The correct four-current for electrons (and positrons), and hence the VP-ESE represented by the $Q$-potential \eqref{CCCQpot},
can only be obtained by the CCC/ETC (see Eqs. (36)--(58) in Ref. \citenum{LiuQED2020}).
Yet, it should be kept in mind that
the CCC is to be applied only when normal ordering with respect to the NES (virtual positrons).
In contrast, the CC (or equivalently the ACR) should still be applied when further normal ordering with respect to the occupied PES\cite{PhysRep},
where charge conjugation is irrelevant. It is in this sense that the $Q$-potential arising from the CCC
should be viewed as an integral part of the Hamiltonian.
In other words, being imposed from the outset, the CCC $\acontraction[0.5ex]{}{a^p}{}{a_q}a^pa_q$ over the vacuum $|0;\tilde{N}\rangle$
should be interpreted as an integral part of the normal-ordering process.

\section{Renormalized Energy}
The renormalized energy of a system of $N$ electrons and zero positrons can readily be calculated as\cite{PCCPNES}
\begin{align}
E&=\langle \Psi(N;\tilde{N})|H^{\mathrm{QED}}_n|\Psi(N;\tilde{N})\rangle-\langle\Psi(0;\tilde{N})|H^{\mathrm{QED}}_n|\Psi(0;\tilde{N})\rangle\label{PhysEn}\\
&=\langle \Psi(N;\tilde{N})|\mathcal{H}|\Psi(N;\tilde{N})\rangle-\langle\Psi(0;\tilde{N})|\mathcal{H}|\Psi(0;\tilde{N})\rangle,\label{PhysEnu}
\end{align}
provided that the CCC \eqref{CCC} is followed when going from Eq. \eqref{PhysEnu} to \eqref{PhysEn}.
Note that the (normalized) wave function $|\Psi(0;\tilde{N})\rangle$ represents here a polarizable vacuum, with $|0;\tilde{N}\rangle$ as its zeroth order.
The same energy can also be calculated as\cite{eQED}
\begin{align}
E&=\langle N;\tilde{N}|H^{\mathrm{QED}}_n|\bar{\Psi}(N;\tilde{N})\rangle,\quad |\bar{\Psi}(N;\tilde{N})\rangle=
\Omega_n|N;\tilde{N}\rangle, \label{MBPT}
\end{align}
or as\cite{PhysRep}
\begin{align}
E&=      \langle\bar{\Psi}(N;\tilde{N})| H^{\mathrm{QED}}_n|\bar{\Psi}(N;\tilde{N})\rangle \label{ExpectEn}\\
 &\equiv \langle\bar{\Psi}(N;\tilde{N})| \mathcal{H}  |\bar{\Psi}(N;\tilde{N})\rangle-C_n. \label{ExpectEnu}
\end{align}
Eqs. \eqref{MBPT} and \eqref{ExpectEn}/\eqref{ExpectEnu} differ only in the normalization (intermediate vs. unitary normalization).
Again, the CCC should be followed when going from Eq. \eqref{ExpectEnu} to \eqref{ExpectEn}. Different from the wave function $\Psi(N;\tilde{N})$
in Eq. \eqref{PhysEn}, the wave function $|\bar{\Psi}(N;\tilde{N})\rangle$ in Eq. \eqref{MBPT}/\eqref{ExpectEn}
is constructed through the action of a normal-ordered wave operator $\Omega_n$ on the non-interacting reference $|N;\tilde{N}\rangle$.
As such, a polarizable vacuum is not needed here (for detailed derivations, see Section IIB.2 and Appendix A in Ref. \citenum{eQED}).
It has been shown\cite{eQED} in detail that the energy calculated by Eq. \eqref{PhysEnu}/\eqref{MBPT}/\eqref{ExpectEn}/\eqref{ExpectEnu} is
in termwise agreement with that by the $\hat{\mathcal{S}}$-matrix (scattering matrix) formulation of QED.
With this more transparent presentation (as compared to Eq. (27) in Ref. \citenum{PhysRep} or Eqs. (48) and (49) in Ref. \citenum{CommentQED}),
it should be clear that
the remark (on Eq. \eqref{ExpectEnu}) that ''the terms subtracted from the referenced Hamiltonians
are single Slater determinants and cannot remove total energy divergence caused by the generalized
electron correlation''\cite{JapQED} does not hold true.

The above energy includes contribution of NESs (virtual positrons) to the correlation of $N$ electrons,
a kind of \emph{derived} QED effect\cite{eQED}. In contrast, what were reported in Ref. \citenum{JapQED} are actually relativistic configuration interaction and
many-body perturbation theories of composite systems of $N$ electrons and $\tilde{M}$ (real) positrons governed by
the instantaneous Coulomb interaction. For instance,
Eqs. (B1), (B4), and B(6) in Ref. \citenum{RespCommentQED} all reduce to the corresponding no-pair Dirac-Coulomb energies
for $N$ electrons by setting $\tilde{M}$ to zero.
Speaking of ``QED Hamiltonian'' in the absence of the VP-ESE (\emph{genuine} QED effect\cite{eQED}) is hardly meaningful.
Likewise, their ``QED-based'' Dirac-Hartree-Fock (DHF) theory\cite{JapQED} is also merely
a relativistic but non-QED mean-field theory\cite{DyallPositron} of $N$ electrons and $\tilde{M}$ positrons.
The genuine QED mean-field theories for electrons only and for both electrons and positrons
were presented in Ref. \citenum{PhysRep} and Refs. \cite{LiuQED2020,eQEDBook2017}, respectively.
In such genuine QED mean-field theories, the PES and NES, whether occupied or not,  are all coupled, thereby fundamentally different
from the relativistic but non-QED counterparts.

\section{Conclusion and outlook}
The many-electron relativistic QED Hamiltonian\cite{eQED} is reformulated as equal average of the Hamiltonian
second-quantized with the electronic Dirac fields and normal-ordered with respect to the filled Dirac sea of electrons and that
second-quantized with the positronic Dirac fields and normal-ordered with respect to the filled Dirac sea of positrons.
This not only renders the introduction of the charge-conjugated contraction
of fermion operators more transparent, but also help clear up the previous misunderstandings. 
Once the effective one-body $Q$ potential responsible for the genuine QED effect (vacuum polarization and electron
self-energy) is represented by a model potential\cite{Uehling1935,Pyykko-SE2003,FG-SE2005,DyallModelQ,ShabaevModelSE,ShabaevModelSEcode2018,Malyshev2022model}, along with the efficient evaluations of
the frequency-independent\cite{LiXiaoSonDCGint,LiXiaoSonDCBint,LiXiaoSonDCBtransformation} 
and frequency-dependent\cite{fdBint2023} two-electron relativistic integrals, the QED Hamiltonian can readily be put into work, following the previous 
unified construction\cite{UnifiedH} of relativistic Hamiltonians
as well as unified implementation\cite{UnifiedWF,Metawave} and parallelization\cite{UnifiedMPI} of size-extensive
and size-consistent wave function methods\cite{PASPT2,PASPT2New}.
High-precision molecular QED calculations that account for the full QED effects
(vacuum polarization, electron self-energy, retardation, and virtual-pairs) on top of essentially converged
relativistic correlation can then be achieved, so 
as to make the left- and right-hand sides of the equation ``Relativity + Correlation + QED = Experiment''\cite{LiuQED2022} as close as possible.

\section*{Acknowledgement}
This work was supported by the National Natural Science Foundation of China (Grant No. 22373057).

\section*{Data Availability Statement}
Data sharing not applicable to this article as no datasets were generated or analysed during the current study.

\clearpage
\newpage

\bibliographystyle{apsrev4-2}
\bibliography{BDFlib}

\begin{thebibliography}{39}%
\makeatletter
\providecommand \@ifxundefined [1]{%
 \@ifx{#1\undefined}
}%
\providecommand \@ifnum [1]{%
 \ifnum #1\expandafter \@firstoftwo
 \else \expandafter \@secondoftwo
 \fi
}%
\providecommand \@ifx [1]{%
 \ifx #1\expandafter \@firstoftwo
 \else \expandafter \@secondoftwo
 \fi
}%
\providecommand \natexlab [1]{#1}%
\providecommand \enquote  [1]{``#1''}%
\providecommand \bibnamefont  [1]{#1}%
\providecommand \bibfnamefont [1]{#1}%
\providecommand \citenamefont [1]{#1}%
\providecommand \href@noop [0]{\@secondoftwo}%
\providecommand \href [0]{\begingroup \@sanitize@url \@href}%
\providecommand \@href[1]{\@@startlink{#1}\@@href}%
\providecommand \@@href[1]{\endgroup#1\@@endlink}%
\providecommand \@sanitize@url [0]{\catcode `\\12\catcode `\$12\catcode
  `\&12\catcode `\#12\catcode `\^12\catcode `\_12\catcode `\%12\relax}%
\providecommand \@@startlink[1]{}%
\providecommand \@@endlink[0]{}%
\providecommand \url  [0]{\begingroup\@sanitize@url \@url }%
\providecommand \@url [1]{\endgroup\@href {#1}{\urlprefix }}%
\providecommand \urlprefix  [0]{URL }%
\providecommand \Eprint [0]{\href }%
\providecommand \doibase [0]{https://doi.org/}%
\providecommand \selectlanguage [0]{\@gobble}%
\providecommand \bibinfo  [0]{\@secondoftwo}%
\providecommand \bibfield  [0]{\@secondoftwo}%
\providecommand \translation [1]{[#1]}%
\providecommand \BibitemOpen [0]{}%
\providecommand \bibitemStop [0]{}%
\providecommand \bibitemNoStop [0]{.\EOS\space}%
\providecommand \EOS [0]{\spacefactor3000\relax}%
\providecommand \BibitemShut  [1]{\csname bibitem#1\endcsname}%
\let\auto@bib@innerbib\@empty
\bibitem [{\citenamefont {Liu}\ and\ \citenamefont {Lindgren}(2013)}]{eQED}%
  \BibitemOpen
  \bibfield  {author} {\bibinfo {author} {\bibfnamefont {W.}~\bibnamefont
  {Liu}}\ and\ \bibinfo {author} {\bibfnamefont {I.}~\bibnamefont {Lindgren}},\
  }\href@noop {} {\bibfield  {journal} {\bibinfo  {journal} {J. Chem. Phys.}\
  }\textbf {\bibinfo {volume} {139}},\ \bibinfo {pages} {014108} (\bibinfo
  {year} {2013})},\ \bibinfo {note} {(E)\textbf{144}, 049901
  (2016).}\BibitemShut {Stop}%
\bibitem [{\citenamefont {Shabaev}(1993)}]{np-eQED}%
  \BibitemOpen
  \bibfield  {author} {\bibinfo {author} {\bibfnamefont {V.~M.}\ \bibnamefont
  {Shabaev}},\ }\href@noop {} {\bibfield  {journal} {\bibinfo  {journal} {J.
  Phys. B}\ }\textbf {\bibinfo {volume} {26}},\ \bibinfo {pages} {4703}
  (\bibinfo {year} {1993})}\BibitemShut {NoStop}%
\bibitem [{\citenamefont {Liu}(2015)}]{IJQCeQED}%
  \BibitemOpen
  \bibfield  {author} {\bibinfo {author} {\bibfnamefont {W.}~\bibnamefont
  {Liu}},\ }\href@noop {} {\bibfield  {journal} {\bibinfo  {journal} {Int. J.
  Quantum Chem.}\ }\textbf {\bibinfo {volume} {115}},\ \bibinfo {pages} {631}
  (\bibinfo {year} {2015})},\ \bibinfo {note} {(E)\textbf{116}, 971
  (2016).}\BibitemShut {Stop}%
\bibitem [{\citenamefont {Kutzelnigg}(2012)}]{KutzFS}%
  \BibitemOpen
  \bibfield  {author} {\bibinfo {author} {\bibfnamefont {W.}~\bibnamefont
  {Kutzelnigg}},\ }\href@noop {} {\bibfield  {journal} {\bibinfo  {journal}
  {Chem. Phys.}\ }\textbf {\bibinfo {volume} {395}},\ \bibinfo {pages} {16}
  (\bibinfo {year} {2012})}\BibitemShut {NoStop}%
\bibitem [{\citenamefont {Liu}(2014{\natexlab{a}})}]{IJQCrelH}%
  \BibitemOpen
  \bibfield  {author} {\bibinfo {author} {\bibfnamefont {W.}~\bibnamefont
  {Liu}},\ }\href@noop {} {\bibfield  {journal} {\bibinfo  {journal} {Int. J.
  Quantum Chem.}\ }\textbf {\bibinfo {volume} {114}},\ \bibinfo {pages} {983}
  (\bibinfo {year} {2014}{\natexlab{a}})}\BibitemShut {NoStop}%
\bibitem [{\citenamefont {Liu}(2014{\natexlab{b}})}]{PhysRep}%
  \BibitemOpen
  \bibfield  {author} {\bibinfo {author} {\bibfnamefont {W.}~\bibnamefont
  {Liu}},\ }\href@noop {} {\bibfield  {journal} {\bibinfo  {journal} {Phys.
  Rep.}\ }\textbf {\bibinfo {volume} {537}},\ \bibinfo {pages} {59} (\bibinfo
  {year} {2014}{\natexlab{b}})}\BibitemShut {NoStop}%
\bibitem [{\citenamefont {Inoue}\ \emph {et~al.}(2023)\citenamefont {Inoue},
  \citenamefont {Watanabe},\ and\ \citenamefont {Nakano}}]{JapQED}%
  \BibitemOpen
  \bibfield  {author} {\bibinfo {author} {\bibfnamefont {N.}~\bibnamefont
  {Inoue}}, \bibinfo {author} {\bibfnamefont {Y.}~\bibnamefont {Watanabe}},\
  and\ \bibinfo {author} {\bibfnamefont {H.}~\bibnamefont {Nakano}},\
  }\href@noop {} {\bibfield  {journal} {\bibinfo  {journal} {J. Chem. Phys.}\
  }\textbf {\bibinfo {volume} {159}},\ \bibinfo {pages} {054105} (\bibinfo
  {year} {2023})}\BibitemShut {NoStop}%
\bibitem [{\citenamefont {Inoue}\ \emph {et~al.}(2024)\citenamefont {Inoue},
  \citenamefont {Watanabe},\ and\ \citenamefont {Nakano}}]{RespCommentQED}%
  \BibitemOpen
  \bibfield  {author} {\bibinfo {author} {\bibfnamefont {N.}~\bibnamefont
  {Inoue}}, \bibinfo {author} {\bibfnamefont {Y.}~\bibnamefont {Watanabe}},\
  and\ \bibinfo {author} {\bibfnamefont {H.}~\bibnamefont {Nakano}},\
  }\href@noop {} {\bibfield  {journal} {\bibinfo  {journal} {J. Chem. Phys.}\
  }\textbf {\bibinfo {volume} {160}},\ \bibinfo {pages} {187102} (\bibinfo
  {year} {2024})}\BibitemShut {NoStop}%
\bibitem [{\citenamefont {Liu}(2024{\natexlab{a}})}]{CommentQED}%
  \BibitemOpen
  \bibfield  {author} {\bibinfo {author} {\bibfnamefont {W.}~\bibnamefont
  {Liu}},\ }\href@noop {} {\bibfield  {journal} {\bibinfo  {journal} {J. Chem.
  Phys.}\ }\textbf {\bibinfo {volume} {160}},\ \bibinfo {pages} {187101}
  (\bibinfo {year} {2024}{\natexlab{a}})}\BibitemShut {NoStop}%
\bibitem [{\citenamefont {Liu}(2020)}]{LiuQED2020}%
  \BibitemOpen
  \bibfield  {author} {\bibinfo {author} {\bibfnamefont {W.}~\bibnamefont
  {Liu}},\ }\href@noop {} {\bibfield  {journal} {\bibinfo  {journal} {J. Chem.
  Phys.}\ }\textbf {\bibinfo {volume} {152}},\ \bibinfo {pages} {180901}
  (\bibinfo {year} {2020})}\BibitemShut {NoStop}%
\bibitem [{\citenamefont {Ferenc}\ \emph {et~al.}(2025)\citenamefont {Ferenc},
  \citenamefont {Salman},\ and\ \citenamefont {Saue}}]{SaueESE2205}%
  \BibitemOpen
  \bibfield  {author} {\bibinfo {author} {\bibfnamefont {D.}~\bibnamefont
  {Ferenc}}, \bibinfo {author} {\bibfnamefont {M.}~\bibnamefont {Salman}},\
  and\ \bibinfo {author} {\bibfnamefont {T.}~\bibnamefont {Saue}},\ }\href@noop
  {} {\bibfield  {journal} {\bibinfo  {journal} {Phys. Rev. A}\ }\textbf
  {\bibinfo {volume} {111}},\ \bibinfo {pages} {L040802} (\bibinfo {year}
  {2025})}\BibitemShut {NoStop}%
\bibitem [{\citenamefont {Benazzouk}\ \emph {et~al.}(2026)\citenamefont
  {Benazzouk}, \citenamefont {Salman},\ and\ \citenamefont
  {Saue}}]{SaueVP-WK2026}%
  \BibitemOpen
  \bibfield  {author} {\bibinfo {author} {\bibfnamefont {R.}~\bibnamefont
  {Benazzouk}}, \bibinfo {author} {\bibfnamefont {M.}~\bibnamefont {Salman}},\
  and\ \bibinfo {author} {\bibfnamefont {T.}~\bibnamefont {Saue}},\ }\href@noop
  {} {\bibfield  {journal} {\bibinfo  {journal} {Phys. Rev. A}\ }\textbf
  {\bibinfo {volume} {113}},\ \bibinfo {pages} {032813} (\bibinfo {year}
  {2026})}\BibitemShut {NoStop}%
\bibitem [{\citenamefont {Uehling}(1935)}]{Uehling1935}%
  \BibitemOpen
  \bibfield  {author} {\bibinfo {author} {\bibfnamefont {E.~A.}\ \bibnamefont
  {Uehling}},\ }\href@noop {} {\bibfield  {journal} {\bibinfo  {journal} {Phys.
  Rev.}\ }\textbf {\bibinfo {volume} {48}},\ \bibinfo {pages} {55} (\bibinfo
  {year} {1935})}\BibitemShut {NoStop}%
\bibitem [{\citenamefont {Pyykk{\"o}}\ and\ \citenamefont
  {Zhao}(2003)}]{Pyykko-SE2003}%
  \BibitemOpen
  \bibfield  {author} {\bibinfo {author} {\bibfnamefont {P.}~\bibnamefont
  {Pyykk{\"o}}}\ and\ \bibinfo {author} {\bibfnamefont {L.-B.}\ \bibnamefont
  {Zhao}},\ }\href@noop {} {\bibfield  {journal} {\bibinfo  {journal} {J. Phys.
  B: At. Mol. Opt. Phys.}\ }\textbf {\bibinfo {volume} {36}},\ \bibinfo {pages}
  {1469} (\bibinfo {year} {2003})}\BibitemShut {NoStop}%
\bibitem [{\citenamefont {Flambaum}\ and\ \citenamefont
  {Ginges}(2005)}]{FG-SE2005}%
  \BibitemOpen
  \bibfield  {author} {\bibinfo {author} {\bibfnamefont {V.}~\bibnamefont
  {Flambaum}}\ and\ \bibinfo {author} {\bibfnamefont {J.}~\bibnamefont
  {Ginges}},\ }\href@noop {} {\bibfield  {journal} {\bibinfo  {journal} {Phys.
  Rev. A}\ }\textbf {\bibinfo {volume} {72}},\ \bibinfo {pages} {052115}
  (\bibinfo {year} {2005})}\BibitemShut {NoStop}%
\bibitem [{\citenamefont {Dyall}(2013)}]{DyallModelQ}%
  \BibitemOpen
  \bibfield  {author} {\bibinfo {author} {\bibfnamefont {K.~G.}\ \bibnamefont
  {Dyall}},\ }\href@noop {} {\bibfield  {journal} {\bibinfo  {journal} {J.
  Chem. Phys.}\ }\textbf {\bibinfo {volume} {139}},\ \bibinfo {pages} {021103}
  (\bibinfo {year} {2013})}\BibitemShut {NoStop}%
\bibitem [{\citenamefont {Shabaev}\ \emph {et~al.}(2013)\citenamefont
  {Shabaev}, \citenamefont {Tupitsyn},\ and\ \citenamefont
  {Yerokhin}}]{ShabaevModelSE}%
  \BibitemOpen
  \bibfield  {author} {\bibinfo {author} {\bibfnamefont {V.~M.}\ \bibnamefont
  {Shabaev}}, \bibinfo {author} {\bibfnamefont {I.}~\bibnamefont {Tupitsyn}},\
  and\ \bibinfo {author} {\bibfnamefont {V.~A.}\ \bibnamefont {Yerokhin}},\
  }\href@noop {} {\bibfield  {journal} {\bibinfo  {journal} {Phys. Rev. A}\
  }\textbf {\bibinfo {volume} {88}},\ \bibinfo {pages} {012513} (\bibinfo
  {year} {2013})}\BibitemShut {NoStop}%
\bibitem [{\citenamefont {Shabaev}\ \emph {et~al.}(2018)\citenamefont
  {Shabaev}, \citenamefont {Tupitsyn},\ and\ \citenamefont
  {Yerokhin}}]{ShabaevModelSEcode2018}%
  \BibitemOpen
  \bibfield  {author} {\bibinfo {author} {\bibfnamefont {V.~M.}\ \bibnamefont
  {Shabaev}}, \bibinfo {author} {\bibfnamefont {I.}~\bibnamefont {Tupitsyn}},\
  and\ \bibinfo {author} {\bibfnamefont {V.~A.}\ \bibnamefont {Yerokhin}},\
  }\href@noop {} {\bibfield  {journal} {\bibinfo  {journal} {Comput. Phys.
  Commun.}\ }\textbf {\bibinfo {volume} {223}},\ \bibinfo {pages} {69}
  (\bibinfo {year} {2018})}\BibitemShut {NoStop}%
\bibitem [{\citenamefont {Malyshev}\ \emph {et~al.}(2022)\citenamefont
  {Malyshev}, \citenamefont {Glazov}, \citenamefont {Shabaev}, \citenamefont
  {Tupitsyn}, \citenamefont {Yerokhin},\ and\ \citenamefont
  {Zaytsev}}]{Malyshev2022model}%
  \BibitemOpen
  \bibfield  {author} {\bibinfo {author} {\bibfnamefont {A.}~\bibnamefont
  {Malyshev}}, \bibinfo {author} {\bibfnamefont {D.}~\bibnamefont {Glazov}},
  \bibinfo {author} {\bibfnamefont {V.}~\bibnamefont {Shabaev}}, \bibinfo
  {author} {\bibfnamefont {I.}~\bibnamefont {Tupitsyn}}, \bibinfo {author}
  {\bibfnamefont {V.}~\bibnamefont {Yerokhin}},\ and\ \bibinfo {author}
  {\bibfnamefont {V.}~\bibnamefont {Zaytsev}},\ }\href@noop {} {\bibfield
  {journal} {\bibinfo  {journal} {Phys. Rev. A}\ }\textbf {\bibinfo {volume}
  {106}},\ \bibinfo {pages} {012806} (\bibinfo {year} {2022})}\BibitemShut
  {NoStop}%
\bibitem [{\citenamefont {Shepler}\ \emph {et~al.}(2005)\citenamefont
  {Shepler}, \citenamefont {Balabanov},\ and\ \citenamefont
  {Peterson}}]{MO-1eQED1}%
  \BibitemOpen
  \bibfield  {author} {\bibinfo {author} {\bibfnamefont {B.~C.}\ \bibnamefont
  {Shepler}}, \bibinfo {author} {\bibfnamefont {N.~B.}\ \bibnamefont
  {Balabanov}},\ and\ \bibinfo {author} {\bibfnamefont {K.~A.}\ \bibnamefont
  {Peterson}},\ }\href@noop {} {\bibfield  {journal} {\bibinfo  {journal} {J.
  Phys. Chem. A}\ }\textbf {\bibinfo {volume} {109}},\ \bibinfo {pages} {10363}
  (\bibinfo {year} {2005})}\BibitemShut {NoStop}%
\bibitem [{\citenamefont {Shepler}\ \emph {et~al.}(2007)\citenamefont
  {Shepler}, \citenamefont {Balabanov},\ and\ \citenamefont
  {Peterson}}]{MO-1eQED2}%
  \BibitemOpen
  \bibfield  {author} {\bibinfo {author} {\bibfnamefont {B.~C.}\ \bibnamefont
  {Shepler}}, \bibinfo {author} {\bibfnamefont {N.~B.}\ \bibnamefont
  {Balabanov}},\ and\ \bibinfo {author} {\bibfnamefont {K.~A.}\ \bibnamefont
  {Peterson}},\ }\href@noop {} {\bibfield  {journal} {\bibinfo  {journal} {J.
  Chem. Phys.}\ }\textbf {\bibinfo {volume} {127}},\ \bibinfo {pages} {164304}
  (\bibinfo {year} {2007})}\BibitemShut {NoStop}%
\bibitem [{\citenamefont {Peterson}(2015)}]{MO-1eQED3}%
  \BibitemOpen
  \bibfield  {author} {\bibinfo {author} {\bibfnamefont {K.~A.}\ \bibnamefont
  {Peterson}},\ }\href@noop {} {\bibfield  {journal} {\bibinfo  {journal} {J.
  Chem. Phys.}\ }\textbf {\bibinfo {volume} {142}},\ \bibinfo {pages} {074105}
  (\bibinfo {year} {2015})}\BibitemShut {NoStop}%
\bibitem [{\citenamefont {Skripnikov}(2021)}]{MOQED-JCP2021}%
  \BibitemOpen
  \bibfield  {author} {\bibinfo {author} {\bibfnamefont {L.~V.}\ \bibnamefont
  {Skripnikov}},\ }\href@noop {} {\bibfield  {journal} {\bibinfo  {journal} {J.
  Chem. Phys.}\ }\textbf {\bibinfo {volume} {154}},\ \bibinfo {pages} {201101}
  (\bibinfo {year} {2021})}\BibitemShut {NoStop}%
\bibitem [{\citenamefont {Sunaga}\ \emph {et~al.}(2022)\citenamefont {Sunaga},
  \citenamefont {Salman},\ and\ \citenamefont {Saue}}]{Saue4CQED2022}%
  \BibitemOpen
  \bibfield  {author} {\bibinfo {author} {\bibfnamefont {A.}~\bibnamefont
  {Sunaga}}, \bibinfo {author} {\bibfnamefont {M.}~\bibnamefont {Salman}},\
  and\ \bibinfo {author} {\bibfnamefont {T.}~\bibnamefont {Saue}},\ }\href@noop
  {} {\bibfield  {journal} {\bibinfo  {journal} {J. Chem. Phys.}\ }\textbf
  {\bibinfo {volume} {157}},\ \bibinfo {pages} {164101} (\bibinfo {year}
  {2022})}\BibitemShut {NoStop}%
\bibitem [{\citenamefont {Schwinger}(1951)}]{Schwinger1951}%
  \BibitemOpen
  \bibfield  {author} {\bibinfo {author} {\bibfnamefont {J.}~\bibnamefont
  {Schwinger}},\ }\href@noop {} {\bibfield  {journal} {\bibinfo  {journal}
  {Phys. Rev.}\ }\textbf {\bibinfo {volume} {82}},\ \bibinfo {pages} {664}
  (\bibinfo {year} {1951})}\BibitemShut {NoStop}%
\bibitem [{\citenamefont {Liu}(2012)}]{PCCPNES}%
  \BibitemOpen
  \bibfield  {author} {\bibinfo {author} {\bibfnamefont {W.}~\bibnamefont
  {Liu}},\ }\href@noop {} {\bibfield  {journal} {\bibinfo  {journal} {Phys.
  Chem. Chem. Phys.}\ }\textbf {\bibinfo {volume} {14}},\ \bibinfo {pages} {35}
  (\bibinfo {year} {2012})}\BibitemShut {NoStop}%
\bibitem [{\citenamefont {Dyall}(2012)}]{DyallPositron}%
  \BibitemOpen
  \bibfield  {author} {\bibinfo {author} {\bibfnamefont {K.~G.}\ \bibnamefont
  {Dyall}},\ }\href@noop {} {\bibfield  {journal} {\bibinfo  {journal} {Chemi.
  Phys.}\ }\textbf {\bibinfo {volume} {395}},\ \bibinfo {pages} {35} (\bibinfo
  {year} {2012})}\BibitemShut {NoStop}%
\bibitem [{\citenamefont {Liu}(2017)}]{eQEDBook2017}%
  \BibitemOpen
  \bibfield  {author} {\bibinfo {author} {\bibfnamefont {W.}~\bibnamefont
  {Liu}},\ }\bibinfo {title} {With-pair relativistic {Hamiltonians}},\ in\
  \href@noop {} {\emph {\bibinfo {booktitle} {Handbook of Relativistic Quantum
  Chemistry}}},\ \bibinfo {editor} {edited by\ \bibinfo {editor} {\bibfnamefont
  {W.}~\bibnamefont {Liu}}}\ (\bibinfo  {publisher} {Springer-Verlag},\
  \bibinfo {address} {Berlin},\ \bibinfo {year} {2017})\ pp.\ \bibinfo {pages}
  {345--373}\BibitemShut {NoStop}%
\bibitem [{\citenamefont {Sun}\ \emph {et~al.}(2021)\citenamefont {Sun},
  \citenamefont {Stetina}, \citenamefont {Zhang}, \citenamefont {Hu},
  \citenamefont {Valeev}, \citenamefont {Sun},\ and\ \citenamefont
  {Li}}]{LiXiaoSonDCGint}%
  \BibitemOpen
  \bibfield  {author} {\bibinfo {author} {\bibfnamefont {S.}~\bibnamefont
  {Sun}}, \bibinfo {author} {\bibfnamefont {T.~F.}\ \bibnamefont {Stetina}},
  \bibinfo {author} {\bibfnamefont {T.}~\bibnamefont {Zhang}}, \bibinfo
  {author} {\bibfnamefont {H.}~\bibnamefont {Hu}}, \bibinfo {author}
  {\bibfnamefont {E.~F.}\ \bibnamefont {Valeev}}, \bibinfo {author}
  {\bibfnamefont {Q.}~\bibnamefont {Sun}},\ and\ \bibinfo {author}
  {\bibfnamefont {X.}~\bibnamefont {Li}},\ }\href@noop {} {\bibfield  {journal}
  {\bibinfo  {journal} {J. Chem. Theory Comput.}\ }\textbf {\bibinfo {volume}
  {17}},\ \bibinfo {pages} {3388} (\bibinfo {year} {2021})}\BibitemShut
  {NoStop}%
\bibitem [{\citenamefont {Sun}\ \emph {et~al.}(2022)\citenamefont {Sun},
  \citenamefont {Ehrman}, \citenamefont {Sun},\ and\ \citenamefont
  {Li}}]{LiXiaoSonDCBint}%
  \BibitemOpen
  \bibfield  {author} {\bibinfo {author} {\bibfnamefont {S.}~\bibnamefont
  {Sun}}, \bibinfo {author} {\bibfnamefont {J.}~\bibnamefont {Ehrman}},
  \bibinfo {author} {\bibfnamefont {Q.}~\bibnamefont {Sun}},\ and\ \bibinfo
  {author} {\bibfnamefont {X.}~\bibnamefont {Li}},\ }\href@noop {} {\bibfield
  {journal} {\bibinfo  {journal} {J. Chem. Phys.}\ }\textbf {\bibinfo {volume}
  {157}},\ \bibinfo {pages} {064112} (\bibinfo {year} {2022})}\BibitemShut
  {NoStop}%
\bibitem [{\citenamefont {Oele}\ \emph {et~al.}(2026)\citenamefont {Oele},
  \citenamefont {Majumder}, \citenamefont {Upadhyay}, \citenamefont {Zhang},
  \citenamefont {Beck}, \citenamefont {Shayit}, \citenamefont {Visscher},\ and\
  \citenamefont {Li}}]{LiXiaoSonDCBtransformation}%
  \BibitemOpen
  \bibfield  {author} {\bibinfo {author} {\bibfnamefont {M.}~\bibnamefont
  {Oele}}, \bibinfo {author} {\bibfnamefont {R.}~\bibnamefont {Majumder}},
  \bibinfo {author} {\bibfnamefont {S.}~\bibnamefont {Upadhyay}}, \bibinfo
  {author} {\bibfnamefont {T.}~\bibnamefont {Zhang}}, \bibinfo {author}
  {\bibfnamefont {R.~A.}\ \bibnamefont {Beck}}, \bibinfo {author}
  {\bibfnamefont {A.}~\bibnamefont {Shayit}}, \bibinfo {author} {\bibfnamefont
  {L.}~\bibnamefont {Visscher}},\ and\ \bibinfo {author} {\bibfnamefont
  {X.}~\bibnamefont {Li}},\ }\href@noop {} {\bibfield  {journal} {\bibinfo
  {journal} {arXiv preprint arXiv:2606.04316}\ } (\bibinfo {year}
  {2026})}\BibitemShut {NoStop}%
\bibitem [{\citenamefont {Inoue}\ and\ \citenamefont
  {Nakajima}(2023)}]{fdBint2023}%
  \BibitemOpen
  \bibfield  {author} {\bibinfo {author} {\bibfnamefont {N.}~\bibnamefont
  {Inoue}}\ and\ \bibinfo {author} {\bibfnamefont {T.}~\bibnamefont
  {Nakajima}},\ }\href@noop {} {\bibfield  {journal} {\bibinfo  {journal} {J.
  Comput. Chem.}\ }\textbf {\bibinfo {volume} {44}},\ \bibinfo {pages} {2073}
  (\bibinfo {year} {2023})}\BibitemShut {NoStop}%
\bibitem [{\citenamefont {Liu}(2024{\natexlab{b}})}]{UnifiedH}%
  \BibitemOpen
  \bibfield  {author} {\bibinfo {author} {\bibfnamefont {W.}~\bibnamefont
  {Liu}},\ }\href@noop {} {\bibfield  {journal} {\bibinfo  {journal} {J. Chem.
  Phys.}\ }\textbf {\bibinfo {volume} {160}},\ \bibinfo {pages} {084111}
  (\bibinfo {year} {2024}{\natexlab{b}})}\BibitemShut {NoStop}%
\bibitem [{\citenamefont {Zhang}\ and\ \citenamefont {Liu}(2024)}]{UnifiedWF}%
  \BibitemOpen
  \bibfield  {author} {\bibinfo {author} {\bibfnamefont {N.}~\bibnamefont
  {Zhang}}\ and\ \bibinfo {author} {\bibfnamefont {W.}~\bibnamefont {Liu}},\
  }\href@noop {} {\bibfield  {journal} {\bibinfo  {journal} {J. Chem. Theory
  Comput.}\ }\textbf {\bibinfo {volume} {20}},\ \bibinfo {pages} {9003}
  (\bibinfo {year} {2024})}\BibitemShut {NoStop}%
\bibitem [{\citenamefont {Zhang}\ \emph {et~al.}(2025)\citenamefont {Zhang},
  \citenamefont {Wang},\ and\ \citenamefont {Liu}}]{Metawave}%
  \BibitemOpen
  \bibfield  {author} {\bibinfo {author} {\bibfnamefont {N.}~\bibnamefont
  {Zhang}}, \bibinfo {author} {\bibfnamefont {Q.}~\bibnamefont {Wang}},\ and\
  \bibinfo {author} {\bibfnamefont {W.}~\bibnamefont {Liu}},\ }\href@noop {}
  {\bibfield  {journal} {\bibinfo  {journal} {J. Phys. Chem. A}\ }\textbf
  {\bibinfo {volume} {129}},\ \bibinfo {pages} {5170} (\bibinfo {year}
  {2025})}\BibitemShut {NoStop}%
\bibitem [{\citenamefont {Wang}\ \emph {et~al.}(2026)\citenamefont {Wang},
  \citenamefont {Zhang},\ and\ \citenamefont {Liu}}]{UnifiedMPI}%
  \BibitemOpen
  \bibfield  {author} {\bibinfo {author} {\bibfnamefont {Q.}~\bibnamefont
  {Wang}}, \bibinfo {author} {\bibfnamefont {N.}~\bibnamefont {Zhang}},\ and\
  \bibinfo {author} {\bibfnamefont {W.}~\bibnamefont {Liu}},\ }\href@noop {}
  {\bibfield  {journal} {\bibinfo  {journal} {J. Chem. Theory Comput.}\
  }\textbf {\bibinfo {volume} {22}},\ \bibinfo {pages} {3499} (\bibinfo {year}
  {2026})}\BibitemShut {NoStop}%
\bibitem [{\citenamefont {Liu}\ \emph {et~al.}()\citenamefont {Liu},
  \citenamefont {Zhang},\ and\ \citenamefont {Liu}}]{PASPT2}%
  \BibitemOpen
  \bibfield  {author} {\bibinfo {author} {\bibfnamefont {C.}~\bibnamefont
  {Liu}}, \bibinfo {author} {\bibfnamefont {N.}~\bibnamefont {Zhang}},\ and\
  \bibinfo {author} {\bibfnamefont {W.}~\bibnamefont {Liu}},\ }\bibfield
  {journal} {\bibinfo  {journal} {Precis. Chem. (2026,
  DOI:10.1021/prechem.5c00408)}\ }\href
  {https://doi.org/10.1021/prechem.5c00408}
  {10.1021/prechem.5c00408}\BibitemShut {NoStop}%
\bibitem [{\citenamefont {Liu}\ \emph {et~al.}(2026)\citenamefont {Liu},
  \citenamefont {Zhang},\ and\ \citenamefont {Liu}}]{PASPT2New}%
  \BibitemOpen
  \bibfield  {author} {\bibinfo {author} {\bibfnamefont {C.}~\bibnamefont
  {Liu}}, \bibinfo {author} {\bibfnamefont {N.}~\bibnamefont {Zhang}},\ and\
  \bibinfo {author} {\bibfnamefont {W.}~\bibnamefont {Liu}},\ }\href@noop {}
  {\bibfield  {journal} {\bibinfo  {journal} {arXiv preprint arXiv:2607.22152}\
  } (\bibinfo {year} {2026})}\BibitemShut {NoStop}%
\bibitem [{\citenamefont {Liu}(2023)}]{LiuQED2022}%
  \BibitemOpen
  \bibfield  {author} {\bibinfo {author} {\bibfnamefont {W.}~\bibnamefont
  {Liu}},\ }\href@noop {} {\bibfield  {journal} {\bibinfo  {journal} {WIREs
  Comput. Mol. Sci.}\ }\textbf {\bibinfo {volume} {13}},\ \bibinfo {pages}
  {e1652} (\bibinfo {year} {2023})}\BibitemShut {NoStop}%
\end{thebibliography}%

\end{document}